\documentclass[prb,aps,twocolumn,showpacs]{revtex4}
\begin{document}

\title{Correlation Effects and the High-Frequency Spin Susceptibility of an Electron Liquid: Exact Limits}
\author{G. S. Atwal}
\author{N. W. Ashcroft}
\affiliation{Cornell Center for Materials Research, and the Laboratory of
Atomic and Solid State Physics, Cornell University, Ithaca, New York 14853-2501}
\date{\today}

\newcommand{\nid}{\noindent}
\newcommand{\ds}{\displaystyle}
\newcommand{\ha}{\widehat}
\newcommand{\hb}{\hbar}
\newcommand{\si}{\sigma}
\newcommand{\la}{\lambda}
\newcommand{\ep}{\epsilon}
\newcommand{\cd}{\cdot}
\newcommand{\de}{\delta}
\newcommand{\na}{\nabla}
\newcommand{\om}{\omega}
\newcommand{\Om}{\Omega}
\newcommand{\da}{\dagger}
\newcommand{\lan}{\langle}
\newcommand{\ran}{\rangle}
\newcommand{\pder}[2]{\frac{\partial{#1}}{\partial {#2}}}
\newcommand{\cder}[1]{\frac{D {#1}}{Dt}}
\newcommand{\pdder}[2]{\frac{{\partial}^2{#1}}{\partial {#2}^2}}
\newcommand{\bearraynn}{\begin{eqnarray}}
\newcommand{\eearraynn}{\end{eqnarray}}
\newcommand{\benn}{\begin{equation}}
\newcommand{\eenn}{\end{equation}}
\newcommand{\eq}[1]{{Eq.~(\ref{#1})}}
\newcommand{\eqs}[2]{{Eqs.~(\ref{#1}--\ref{#2})}}
\newcommand{\lab}{\label}
\newcommand{\frd}[2]{\ds \frac{{#1}}{{#2}}}
\newcommand{\cn}[2]{\chi_{#1}({\bf q},{#2})}
\newcommand{\wit}{\tilde{\om}}
\newcommand{\ppq}{{\bf p}\! + \! \frac{1}{2}{\bf q}}
\newcommand{\pmq}{{\bf p}\! - \! \frac{1}{2}{\bf q}}
\newcommand{\dep}{\lan \ppq | \ha{\de \rho} | \pmq \ran}
\newcommand{\De}{\Delta}
\newcommand{\wmpq}{\frac{\om m}{p\, q}}
\newcommand{\pqwm}{\frac{p\, q}{\om m}}
\newcommand{\pfqwm}{\frac{p_F q}{\om m}}
\newcommand{\wmpfq}{\frac{\om m}{p_F q}}
\newcommand{\gud}{g_{\uparrow \downarrow}}


\begin{abstract}
Spin correlations in an interacting electron liquid are studied in the
high-frequency limit and in both two and three dimensions. The
third-moment sum rule is evaluated and used to derive exact
limiting forms (at both long- and short-wavelengths) for the spin-antisymmetric local-field factor, $\lim_{\om \to \infty}G_-({\bf q, \om})$. In two dimensions $\lim_{\om \to \infty}G_-({\bf q, \om})$ is found to diverge as $1/q$ at long wavelengths, and the spin-antisymmetric exchange-correlation kernel of time-dependent spin density functional theory diverges as $1/q^2$ in both two and three dimensions. These signal a failure of the local-density approximation, one that can be redressed by alternative approaches.
\end{abstract}

\maketitle

\section{Introduction}
The detailed incorporation of exchange and correlation (xc) into the dynamic
response of an interacting electron liquid remains a challenge to theory. The mean field approach of the random phase
approximation (RPA), which neglects xc effects, becomes increasingly
inadequate as both the dimensionality and density of electrons is
lowered. Sum rules provide exact constraints on the response of an
electron liquid and thus are extremely useful as tests of any scheme
that purports to include xc effects. Alternatively they can be used to
constrain particular approximations such as those embodied in local field factors
$G_{\pm}({\bf q},\om)$. The static behavior of both the spin-symmetric
($+$) and spin-antisymmetric ($-$) local field factors has been well
studied \cite{Iwa84} and various fitting formulae have been proposed
that conform to the known exact constraints \cite{DavPol011,DavPol012,CorSol98}. Their dynamic forms 
have attracted a great deal of recent attention since they exhibit far richer
behavior than their static counterparts. 

However, up until now, most dynamical studies have
focussed on the spin-symmetric case; in this paper we
report on a derivation of the exact behavior of the dynamic
spin-antisymmetric response in the high-$\om$ limit within the linear
response scheme. In particular, the first and third-moment sum rules are derived (Sec.\ref{se:mome}) and then used in Sec.\ref{se:loca} to determine the exact
asymptotic behaviour of $\lim_{\om \to \infty}G_-({\bf q},\om)$ in the low- and high-$q$
limit. The xc kernel in time-dependent density functional theory is closely related to these local field factors and in Sec.\ref{se:disc} it will be
shown that the high-frequency spin-antisymmetric xc kernel must then {\it diverge} in the long wavelength
limit. This divergence demonstrates that the dynamical xc potential in time-dependent spin-density functional theory (TDSDFT), as applied to a homogeneous system, must be a {\it nonlocal} functional of the local density.

\section{Moment Sum Rules}
\label{se:mome}
To describe the formalism needed to derive the moment sum rules of an
electron liquid we will first need the electron particle-hole operator $\hat{\rho}_{{\bf k}\si}({\bf q})$, defined as
\benn
\hat{\rho}_{{\bf k} \si}({\bf q})=c_{{\bf k} \si}^{\da} c_{{\bf k}+{\bf q}, \si},
\eenn
\nid where  $c_{{\bf q} \si}^{\da}$ and $c_{{\bf q} \si}$  are the
fermionic creation and annihilation operators, respectively, and $\si$
denotes the spin projection. The standard one-particle density
operators are then
\benn \hat{\rho}_{\si}({\bf q})=\sum_{\bf k} \hat{\rho}_{{\bf k}
\si}({\bf q}), \hspace{1cm} \hat{\rho}({\bf q})=\sum_{{\bf k} \si} \hat{\rho}_{{\bf k} \si}({\bf q}),
\eenn
\nid and, for the spin equivalent,
\benn
\hat{s}({\bf q})= \sum_{{\bf k}\si} \eta_{\si} \rho_{{\bf k} \si} ({\bf q}), \hspace{1cm} \eta_{\si}= \left\{ \begin{array}{l}
+1 \hspace{2mm} \hbox{for} \hspace{2mm} \si=\uparrow, \\
-1 \hspace{2mm} \hbox{for} \hspace{2mm} \si=\downarrow.
\end{array} \right.
\eenn
In the thermodynamic limit the Hamiltonian for a $d$-dimensional system of $N$ electrons immersed in a uniform neutralizing background of volume $\Om$ is given by ($\hbar=1$),
\benn
H=\sum_{{\bf q} \si} \om_0({\bf q}) c_{{\bf q} \si}^{\da} c_{{\bf q}
\si} + \frac{N}{2 \Om} \sum_{{\bf q} \neq 0} V({\bf q}) [ \frac{1}{N} \hat{\rho}({\bf q}) \hat{\rho} (-{\bf q}) - 1] \lab{Hami} \eenn
\nid where $\om_0({\bf q})=q^2/2m$, and  $V({\bf q})$ is the Fourier
transformed Coulomb potential, i.e. $V({\bf q})=4\pi e^2/q^2$ for $d=3$,
and $V({\bf q})=2\pi e^2/q$ for $d=2$.
From linear response theory the retarded charge-density
$\chi_C$ and spin-density $\chi_S$ responses are given by
\benn
\chi_C({\bf q},t)=\frac{i}{\Om} \lan [ \hat{\rho}({\bf q},t), \hat{\rho}(-{\bf q},0)] \ran \theta(t),
\eenn
\nid and
\benn
\chi_S({\bf q},t)=\frac{i}{\Om} \lan [ \hat{s}({\bf q},t), \hat{s}(-{\bf q},0)]
\ran \theta(t), \lab{chis}
\eenn
\nid where $\theta(t)$ is the Heaviside step function and $\lan \ran$
denotes averaging over the equilibrium ensemble as specified by the
Hamiltonian in \eq{Hami}. Note that to obtain a physical spin-spin response from $\chi_S({\bf q},t)$ in \eq{chis} a factor $g^2 \mu_B^2/4$ is required where $g$ is the
gyromagnetic ratio and $\mu_B$ is the Bohr magneton. 

As usual, the spectral representation of $\chi({\bf q}, \om)$, the Fourier transform of $\chi({\bf q}, t)$, can be obtained from
\benn \chi({\bf q}, \om)=\int_{-\infty}^{\infty} \frac{d \om'}{\pi} \frac{{\mathrm{Im}} \chi({{\bf q}, \om'})}{\om'-\om-i0^+}. \eenn
\nid Analyticity of $\chi_S({\bf q}, \om)$ in the upper half of the complex-$\om$ plane requires that ${\mathrm{Im}} \chi_S({\bf q},\om)$ be an odd function of $\om$ and so, in the the high-$\om$ limit, $\chi_S({\bf q}, \om)$ expands as
\benn
\lim_{\om \to \infty} \chi_S({\bf q},\om)=-\sum_{l=1}^{\infty} \frac{\lan \om^{(2l-1)} \ran}{\om^{2l}}, \lab{chix}
\eenn
\nid where the moments $\lan \om^{(l)} \ran$ are given by
\bearraynn
\lan \om^{(l)} \ran &\equiv&   \int_{-\infty}^{\infty} \frac{ d\om}{\pi} \om^{l} {\mathrm{Im}} \chi_S({\bf q},\om), \nonumber \\
&=& \frac{1}{\Om} \left\lan \left[ \left(i \pder{}{t} \right)^l \hat{s}({\bf q},t), \hat{s}(-{\bf q},0) \right]  \right\ran_{t=0}.
\eearraynn
The first moment yields the well known $f$-sum rule, 
\benn
\lan \om^{(1)} \ran = 2 n \om_0({\bf q}) \lab{first}
\eenn
\nid where $n\equiv N/\Om$ is the macroscopic homogeneous particle density. The derivation of the $f$-sum rule invokes number conservation via the continuity equation,
\benn
\pder{\rho_{\si}({\bf q},t)}{t}=-i {\bf q} \cd {\bf j}_{\si}({\bf q},t)
\eenn
\nid where ${\bf j_{\si}}({\bf q},t)$ is the current density fluctuation operator for particles with spin $\si$
\benn
{\bf j}_{\si}({\bf q},t)=\frac{1}{m} \sum_{{\bf k}} ({\bf k}+\frac{1}{2}{\bf q}) \rho_{{\bf k} \si}({\bf q},t).
\eenn
Thus number density is {\it separately} conserved for electrons with
spin up and down since there is no term in the equilibrium Hamiltonian
\eq{Hami} which can flip spin. That correlation effects play no role
in the first moment may be understood by noting that in the high-$\om$
limit electrons which interact via a velocity-independent potential
cannot be influenced by the effects of others when the time interval
is infinitesimal. 

The third-moment sum rule \cite{Puf65}, after straightforward but
lengthy calculations, gives \cite{mythesis,SinPat74}
\newcommand{\wpq}[1]{\om_{{\mathrm{p}}}^{#1}({\bf q})}
\benn 
\lan \om^{(3)} \ran =2 n \om_0({\bf q})[ \om_0^2({\bf q}) + \om_0({\bf q})(12/d) \lan E_{KE} \ran  + I_d({\bf q})] \lab{third}
\eenn
\nid where $\lan E_{KE} \ran$ is the average kinetic energy per electron in the interacting system and $I_d({\bf q})$ is given by
\benn
I_d({\bf q})=\frac{1}{m} \sum_{{\bf k} \neq 0} V({\bf k}) (\hat{\bf q} \cd {\bf k})^2 [ \tilde{S}({\bf q} - {\bf k})- S({\bf k}) ],
\eenn 
\nid where for ${\bf q} \neq 0$
\bearraynn \tilde{S}({\bf q})&=&N^{-1} \lan \hat{s}({\bf q}) \hat{s}(-{\bf q}) \ran, \\
S({\bf q})&=&N^{-1} \lan \hat{\rho}({\bf q}) \hat{\rho}(-{\bf q}) \ran, \eearraynn
define the static magnetic and usual structure factors, respectively. For completeness, we express the structure factors in terms of the spin-resolved pair correlation functions,
\benn
\tilde{S}(k)-1=\frac{n}{2}\int d^2r \left(g_{\uparrow \uparrow}(r)
-g_{\uparrow \downarrow}(r) \right) \exp(-i {\bf k} \cd {\bf
r}),
\eenn
\nid and
\benn
{S}(k)-1=\frac{n}{2}\int d^2r \left(g_{\uparrow \uparrow}(r)
+g_{\uparrow \downarrow}(r) -2\right) \exp(-i {\bf k} \cd {\bf
r}).
\eenn
\nid where 
$g_{\uparrow \uparrow}(r)$ and $g_{\uparrow \downarrow}(r)$ are the spin-parallel and antiparallel contributions, respectively, to the electron pair correlation function $g(r)$, i.e.

\benn g(r)=\frac{1}{2}[g_{\uparrow \uparrow}(r) + \gud(r)].\eenn

  It has been noted \cite{GooSjo73} that the third-moment sum rule differs for the charge and spin density response, in contrast to the first moment sum rule which is obeyed by both $\chi_C$ and $\chi_S$. The physical root of this qualitative difference can be inferred from the continuity equation for longitudinal spin-current, used in the derivation of \eq{third}, namely
\begin{widetext}
\benn
i\pder{}{t} {\bf q} \cd {\bf j}_{\si}({\bf q},t)=\sum_{\bf k} [\om_0({\bf q}+{\bf k})-\om_0({\bf k})]^2 \rho_{{\bf k} \si}({\bf q},t) + \frac{1}{\Omega} \sum_{{\bf k}{\bf p}}V({\bf p})\frac{{\bf q}\cd{\bf p}}{m}c^{\da}_{{\bf k} \si}(t) \rho({\bf p},t)c_{{\bf k}-{\bf p}+{\bf q} \si}(t).
\eenn
\end{widetext}
\nid Hence, it can be shown that the presence of the inter-electron potential
allows a transfer of momentum density across spin-up and spin-down
electrons. Thus, although the total momentum density must be
conserved, the momentum density of each spin species is not
\cite{AmiVig00}. This is why the third-moment sum rule, essentially
an expression of momentum and number density conservation, must differ for charge and spin density susceptibilities.

An expression for $I_d({\bf q})$ has been known for some time for
three-dimensional systems \cite{SinPat74}, and here we just record the limiting expressions, 
\benn
I_3({\bf q} \to 0)=\frac{\wpq{2}}{3}[1-g_{\uparrow \downarrow}(0)], \lab{3lolimit}
\eenn
\nid and
\benn
I_3({\bf q} \to \infty)=\frac{\wpq{2}}{3}[1-2g_{\uparrow \downarrow}(0)], \lab{3hilimit}
\eenn
\nid where $\om_{\mathrm{p}}({\bf q})=[2 n  \om_0({\bf q}) V({\bf q})]^{1/2}$ is the plasma frequency. Note that in \eqs{3lolimit}{3hilimit} we have applied \cite{GooSjo73} $g_{\uparrow
\uparrow}(0)=0$, and that the high-${\bf q}$ limit is to be taken such that $0 \ll \om({\bf q}) \ll \om$.  

 In two dimensions we find
\benn
I_2({\bf q})=\frac{e^2}{2 m}  \int_0^\infty dk \, k \left\{ R({\bf q},{\bf k})  [\tilde{S}(k)-1] - k [S(k)-1] \right\} 
\eenn
with
\benn
R({\bf q},{\bf k})= \int_{0}^{2 \pi} \frac{d\theta}{\pi} \frac{q^2 +k^2 \cos^2 \theta + 2kq \cos \theta}{(q^2+k^2+2kq \cos \theta)^{1/2}}.
\eenn
\nid The limiting behavior is then
\bearraynn
I_2({\bf q} \to 0)&=&\frac{e^2}{2m} \int_{0}^{\infty} dk \, k^2 [\tilde{S}(k) - S(k)], \\
I_2({\bf q} \to \infty)&=&-\frac{1}{2} \wpq{2} \gud(0).
\eearraynn

\section{Local-Field Factors}
\label{se:loca}
\newcommand{\lind}{\overline{\chi}_0({\bf q},\om)}
Another way to express the correlation effects in the spin response is
to introduce the notion of a local-field factor, $G_-({\bf q},\om)$,
which modifies the effective field felt by a single electron, as
expressed by
\benn \chi_S({\bf q},\om)=-\frac{\overline{\chi}_0({\bf q},\om)}{1+V({\bf q}) G_- ({\bf q},\om) \lind}. \lab{defG} \eenn
\nid Here $\lind$ is the {\it modified} Lindhard function for the
 interacting system, i.e. where {\it exact} interacting occupation numbers, 
$\overline{n}({\bf k})$, are used, namely
\benn
\lind=\frac{1}{\Omega}\sum_{\bf k} \frac{\overline{n}({\bf k})-
\overline{n}({\bf k}+{\bf q})}{\om + \om_0({\bf k}) - \om_0({\bf
k}+{\bf q})+i0^+}. \lab{mlind} \eenn
\nid The $\overline{n}({\bf k})$ therefore replace the usual non-interacting fermion occupation number, $n({\bf k})$, in the original Lindhard function, $\chi_0({\bf q},\om)$. Since the exact occupation numbers are generally unknown we follow Ref.\
\onlinecite{RicAsh94} and introduce a separate local factor, $G_n$ to
write the modified Lindhard function in terms of the original Lindhard
function $\chi_0$, i.e.
\benn
\lind=\frac{\chi_0({\bf q},{\om})}{1+V({\bf q}) G_n({\bf q},\om) \chi_0({\bf q},\om)}. \eenn
\nid Thus we have
\benn
\chi_S({\bf q},\om)=-\frac{\chi_0 ({\bf q},\om)}{1+V({\bf q}) \overline{G}_-({\bf q},\om) \chi_0({\bf q},\om)}, \lab{chif} \eenn
\nid where $\overline{G}_-({\bf q},\om)=G_-({\bf q},\om) + G_n({\bf q},\om)$.

The third-moment sum rule can again be utilised to determine the high-$\om$ limit of the local-field factors, i.e. $\lim_{{\om} \to \infty} G({\bf q},\om)=G^{(\infty)}({\bf q})$. First we will require the high-$\om$ expansion of the Lindhard function, i.e.
\begin{widetext}
\benn
\lim_{{\om} \to \infty}\chi_0({\bf q},\om)=\frac{2n \om_0({\bf
q})}{\om^2}+\frac{2n\om_0^2({\bf q})}{\om^4}\left[ \om_0({\bf
q})+\frac{12}{d} \lan E_{KE} \ran_0 \right]+
O\left(\frac{1}{\om^6} \right) \lab{highlind}
\eenn
\nid where $\lan E_{KE} \ran_0$ is the average kinetic energy per
electron in a non-interacting system. After insertion of \eq{highlind} into the
high-$\om$ expansion of the spin-density response \eq{chif} we observe
that manifestations of the local field factor appear at $O(1/\om^4)$, i.e. 
\benn
\lim_{{\om} \to \infty}\chi_S({\bf q},\om)=-\frac{2 n \om_0({\bf q})}{\om^2}-\frac{2 n \om_0^2({\bf q})}{\om^4} \left[ \om_0({\bf q})+\frac{12}{d} \lan E_{KE} \ran_0 - V({\bf q}) 2n \overline{G}_{-}^{(\infty)}({\bf q}) \right]+O\left(\frac{1}{\om^6}\right).
\eenn
\end{widetext}
\nid By comparison with the first- and third-moment sum rules, \eq{first} and \eq{third}, we infer that limiting form of the local-field factor must be given by
\benn
\overline{G}_{-}^{(\infty)}({\bf q})=-\frac{I_d({\bf q})}{\wpq{2}} + \frac{6}{d \, n V({\bf q})} (\lan E_{KE} \ran_0 -\lan E_{KE} \ran).  \eenn
The kinetic part of the above expression gives a diverging
contribution in the large $q$ limit which, previous authors \cite{GooSjo73} have
argued, must be absorbed by a proper treatment of the self-energy of
the proper polarizability function, i.e.  the modified Lindhard
function. The introduction of the additional local-field factor, $G_n$, allows us to demonstrate this explicitly, and indeed a high-${\om}$ expansion of \eq{mlind} together with the definition of the local field-factor, \eq{chif}, gives
\benn
G_n^{(\infty)}({\bf q})= \frac{6}{d \, n V({\bf q})} (\lan E_{KE} \ran_0 -\lan E_{KE} \ran), \eenn
\nid leaving the desired result,
\benn
G_-^{(\infty)}({\bf q})= -\frac{I_d({\bf q})}{\wpq{2}}.
\eenn
\nid This result can also be derived from the work of Zhu and Overhauser\cite{ZhuOve84} based on the equation-of-motion approach\cite{Nik74}. In three dimensions the limiting forms are
\bearraynn
G_-^{(\infty)}({\bf q} \to 0)&=&\frac{1}{3}[2 g(0)-1], \lab{3d} \\
G_-^{(\infty)}({\bf q} \to \infty)&=&\frac{1}{3}[4 g(0) -1],
\eearraynn
\nid Note that if the negative sign of the local-field factor at high $\om$, by \eq{3d}, persists at finite $\om$ then we have the possibility that a pole in $\chi_S({\bf q},\om)$ exists \cite{UtsIch83} for real $\om$ and sufficiently large $r_s$. Physically, this collective mode corresponds to long-wavelength spin-wave modes, undamped and oscillating at frequencies given by the solution to
\benn
1+V({\bf q})G_-({\bf q},\om)\overline{\chi}_0({\bf q},\om)=0
\eenn
In two dimensions the limiting forms of the high-frequency local-field factors are
\bearraynn
G_-^{(\infty)}({\bf q} \to 0)&=&\frac{1}{4 \pi n q} \int_{0}^{\infty}
dk \, k^2 [\tilde{S}(k)-S(k)], \lab{2d} \\
G_-^{(\infty)}({\bf q} \to \infty)&=&g(0). 
\eearraynn

\section{Discussion}
\label{se:disc}
The physical concept behind the introduction of the local-field factor
is identical in spirit to the methodology of DFT
in which xc effects are incorporated in an
effective local field experienced by each electron. In fact, within the frequency-dependent linear response formalism \cite{GroKoh85}, the
unpolarized spin-antisymmetric local-field factor is simply related to the xc energy functional $E_{\rm{xc}}[n]$ via, 
\benn
G_-({\bf q},\om)=\frac{K_{\rm{xc}}^{\uparrow \downarrow}({\bf q},\om)-K_{\rm{xc}}^{\uparrow \uparrow}({\bf q},\om)}{2 V({\bf q})} \equiv - \frac{K_{\rm{xc}}^-({\bf q},\om)}{2 V({\bf q})}  \eenn
\nid where $K_{\rm{xc}}^{\sigma \sigma'}({\bf q},\om)$ is the
Fourier-transform of the xc kernel of TDSDFT, 
\benn K_{\rm{xc}}^{\sigma \sigma'}({\bf r}-{\bf r}',t-t') \equiv \frac{\de^2 E_{\rm{xc}}[n]}{\de n_{\sigma}({\bf r},t) \de n_{\sigma'}({\bf r},t')}. \eenn
From \eq{3d} and \eq{2d} we now note that $\lim_{{\bf q}\to 0} \lim_{\om
\to \infty}K_{\rm{xc}}^-({\bf q},\om)$ diverges as $q^{-2}$ in both
two and three dimensions in notable contradistinction to $K_{\rm{xc}}^+({\bf q},\om)$ which is well-behaved in the same limit. This striking behavior of the spin-antisymmetric
exchange-correlation kernel does not by itself lead to any concern since
the corresponding macroscopic quantities (i.e. $\chi_S$ as given by \eq{chif})
remain well-defined in the low-$q$ and high-$\om$ limit and it
follows that the observable (i.e. induced magnetization) also remains
well-defined. Physically, the divergence in momentum-space implies that $K_{\rm{xc}}^-(r,\om)$, and hence the xc hole, is of infinite extent, i.e. it decays very slowly, especially so in two dimensions. The divergence is then a signature of a transition from a bounded xc kernel to an unbounded one. Thus a strict
local-density approximation (LDA)\cite{GroKoh85} for the xc potential
becomes untenable in time-dependent systems, at least in the
high-frequency regime. In other words, the
dynamical local field cannot be specified by the {\it local} density alone
and other approaches to TDSDFT are required. This divergence is particularly potent given that it occurs even in a homogeneous system, as emphasised by Vignale \cite{VigAPS02} who very recently reported similar conclusions about the dynamic spin response at finite-$\om$.

One possible remedy is via the weighted density approximation
\cite{GunJon78} which retains nonlocality and has been demonstrated to
give more accurate results than functional approximations involving
local-density contributions. Indeed it has been argued \cite{MagAsh87,RapAsh91}
that such an approach should be able to recover at least part of the
the full Van der Waals interaction, the origin of which is clearly due
to dynamical correlation between electrons, and where it is well known that the
conventional LDA fails. Another approach which overcomes the limitations of the LDA but
nonetheless still
permits a local description is current density functional theory
\cite{VigKoh96,VigUll97} where the local current density is required
in addition to the local density to describe time-dependent
inhomogeneous systems.

Finally, we note that many-body calculations based on Hedin's $GW$
approximation \cite{Hed65} take correlations into account directly
without invoking the local mean-field approximations of DFT, and have
recently \cite{GarGod02} been demonstrated to yield accurate
ground-state energies of inhomogeneous systems where LDA fails
drastically. However, such many-body approaches are numerically intensive and it remains to be seen whether they can be
successfully applied to excited states within feasible computational effort.

\begin{acknowledgments}
This work was supported by the NSF under Grant No. DMR-9988576.
\end{acknowledgments}

\bibliography{ref}

\end{document}